\documentclass[reprint, groupedaddress,aps
]{revtex4-2}

\usepackage{hyperref}
\usepackage{algorithm}
\usepackage{algorithmic}
\usepackage{blkarray}
\usepackage{soul}
\usepackage{comment}
\usepackage{graphicx}
\usepackage{dcolumn}
\usepackage{booktabs} 
\usepackage{amsmath,amssymb}
\usepackage{color}
\usepackage{multirow}
\usepackage{array}

\hypersetup{
    colorlinks = true,
    linkcolor = blue,
    citecolor = blue,
    filecolor = blue,
    urlcolor = blue
}

\makeatother
\begin{document}

\preprint{APS/123-QED}

\title{Scalable Connectivity for Ising Machines: Dense to Sparse }%
\author{M Mahmudul Hasan Sajeeb$^{1}$}
\author{Navid Anjum Aadit$^{1}$}
\author{Shuvro Chowdhury$^{1}$}
\author{Tong Wu$^{2}$}
\author{Cesely Smith$^{2}$}
\author{Dhruv Chinmay$^{2}$}
\author{Atharva Raut$^{2}$}
\author{Kerem Y. Camsari$^{1}$}
\author{Corentin Delacour$^{1}$}
  \email{delacour@ucsb.edu}
\author{Tathagata Srimani$^{2}$}
  \email{tsrimani@andrew.cmu.edu}

\affiliation{$^1$Department of Electrical and Computer Engineering, University of California, Santa Barbara, CA, 93106, USA} 

\affiliation{$^2$Department of Electrical and Computer Engineering, Carnegie Mellon University, Pittsburgh, PA, 15213, USA} 

\date{\today}

\begin{abstract}
In recent years, hardware implementations of Ising machines have emerged as a viable alternative to quantum computing for solving hard optimization problems among other applications. Unlike quantum hardware, dense connectivity can be achieved in classical systems. However, we show that dense connectivity leads to severe frequency slowdowns and interconnect congestion scaling unfavorably with system sizes.  As a scalable solution, we propose a systematic sparsification method for dense graphs by introducing copy nodes to limit the number of neighbors per graph node. In addition to solving interconnect congestion, this approach enables constant frequency scaling where all spins in a network can be updated in constant time. On the other hand, sparsification introduces new difficulties, such as constraint-breaking between copied spins and increased convergence times to solve optimization problems, especially if exact ground states are sought. Relaxing the exact solution requirements, we find that the overheads in convergence times are milder. We demonstrate these ideas by designing probabilistic bit Ising machines using ASAP7 (a predictive 7nm FinFET technology model) process design kits as well as Field Programmable Gate Array (FPGA)-based implementations.
Finally, we show how formulating problems in naturally sparse networks (e.g., by invertible logic)  sidesteps challenges introduced by sparsification methods. Our results are applicable to a broad family of Ising machines using different hardware implementations. 
\end{abstract}

\maketitle

\section{Introduction}
\label{Introduction}

Physics-inspired hardware platforms like Ising Machines (IMs) have gained attention for tackling computationally hard problems, leveraging energy minimization principles for combinatorial optimization and probabilistic sampling.
In essence, an Ising machine solves an input problem represented as a graph, either physically or virtually, by constructing a  network of coupled \textit{spins} ($m_i=\pm1$) that evolves to minimize the Ising Hamiltonian:
\begin{equation}
    E=-\sum_{i<j}J_{ij}m_i m_j-\sum_i h_i m_i
    \label{eq:ising_hamiltonian}
\end{equation}
where $J_{ij}$ is the coupling between two spins, and $h_i$ is the spin bias. Many NP-hard problems have been mapped to Eq.~\ref{eq:ising_hamiltonian} using various techniques \cite{lucas2014ising}. As a result, physically realizing Ising machines to implement state-of-the-art probabilistic algorithms hold significant potential to accelerate hard optimization tasks that are intractable at large scales.
Ising machines have been realized using a variety of technologies leveraging distinct physics. These include quantum circuits \cite{johnson2011quantum,king2023quantum, king2024computational}, lasers \cite{mcmahon2016fully,honjo_2021}, memristors \cite{fahimi_2021,jiang_2023}, coupled oscillators \cite{moy20221,Graber_2024}, nanodevices \cite{camsari2019scalable, singh2024cmos},  digital CMOS circuits \cite{aadit2022massively,Hitachi_yamaoka2015,smithson2019efficient, Fujitsu_aramon2019physics} and others.  Most recent Ising machines \cite{lo2023ising, hamerly2018scaling,goto2019combinatorial, aramon2019physics} have emphasized all-to-all connectivity presumably with the motivation of reconfigurability: in an all-to-all graph, \textit{any} problem expressed in the form of Eq.~\ref{eq:ising_hamiltonian} can be programmed onto the hardware. This is in stark contrast with quantum annealers from D-Wave whose cryogenic hardware necessitates sparsity in networks \cite{king2023quantum, king2024computational}.  

\begin{figure*}[t!]
    \centering
    \includegraphics[width=1\linewidth]{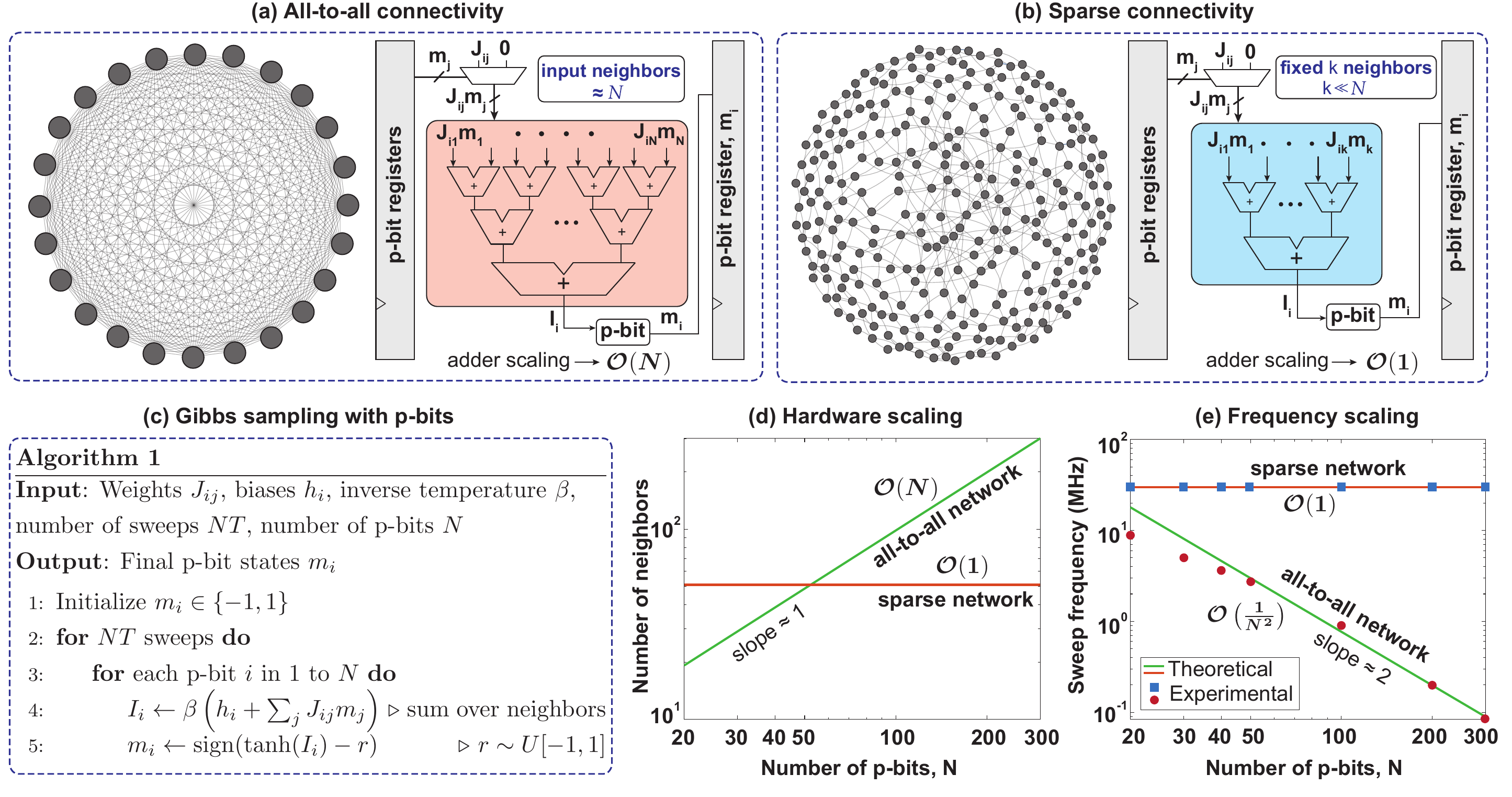}
    \vspace{-20pt}
    \caption{(a-b) All-to-all vs sparse Ising machines with probabilistic bits (p-bit). For all-to-all connectivity, each p-bit requires adding the contribution of $N-1$ neighbors before updating. For sparse connectivity with an average degree $k$, only $k$ neighbors are added before a p-bit updates. (c) Pseudocode for Gibbs sampling \cite{koller2009probabilistic}  for p-bit-based Ising machines. Line 3 shows \textit{sequential} updates and line 4 shows addition over neighbors. (d) The number of neighbors per node for all-to-all and sparse connectivity scale as $\mathcal{O}(N)$ and $\mathcal{O}(1)$, respectively. (e)  All-to-all requires \(N\) clock cycles per Monte Carlo sweep (MCS) due to sequential updates and increasing adder sizes, scaling as \(\mathcal{O}(N^{-2})\). Sparse networks, by parallelizing independent p-bits, maintain constant MCS frequency, matching theoretical predictions from FPGA experiments (see Methods).}
    \label{fig1}
    \vspace{-10pt}
\end{figure*}

As we show in this work, all-to-all connectivity poses severe scaling challenges for Ising machines: the most obvious difficulty is the quadratically growing number of connections which causes routing difficulties. The second, more subtle point is algorithmic: nodes in an Ising machine typically evolve sequentially even if their description is parallel in continuous time \cite{suzuki2013chaotic,lee2025noise}. Node updates are \textit{conditioned} on their neighbors to properly reduce energy or converge to the Boltzmann distribution. As such, all-to-all connectivity requires each spin to receive $\mathcal{O}(N)$ additions from neighboring spins before updates (FIG.~\ref{fig1}a-b). 

The combined effect of dense routing and growing additions per node makes sparse  representations inevitable for scaled implementations. In the context of quantum annealers, comparisons between sparse and all-to-all graph topologies exist \cite{hamerly_2019,venturelli_2015}. However, the limitations of quantum annealers make these comparisons highly specific. Emerging Ising machines enjoy far greater flexibility. Our purpose in this work is to systematically analyze sparse vs. all-to-all network topologies for a broader class of Ising machines by decoupling algorithm, architecture and technology contributions to performance.  While our results focus on p-bit based Ising machines, the connectivity-related trade-offs we study such as reduced update parallelism and the need for scalable embeddings may be applicable to a broader class of Ising machines, including those that use analog or optical summation mechanisms.

This work focuses on \textit{probabilistic-bit} (p-bit) based IMs (p-computers) with spin dynamics (FIG.~\ref{fig1}c) that sample from the Boltzmann distribution \cite{bohm2022noise,chowdhury2023full}. While our examples use the p-bit framework, the conclusions apply broadly to other Ising machines. We propose a systematic sparsification algorithm that transforms dense problems into sparse ones using auxiliary copy nodes, without altering the problem's ground state. However, in practice, sparsification can introduce infeasible solutions due to disagreements among copy nodes, significantly increasing the required Monte Carlo steps to maintain success probabilities. We find that if approximate solutions are acceptable, the overheads are much smaller. We also synthesize all-to-all and sparse Ising machines using the ASAP7 \cite{clark2016asap7} process design kit (PDK), confirming scaling laws in area and frequency for both architectures. Finally, we highlight the advantages of alternative, \textit{natively sparse} problem formulations over sparsifying dense graphs.

\section{Scaling Features of All-to-all vs. Sparse}

FIG.~\ref{fig1} sets the stage for all-to-all vs. sparse connectivity in p-bit based Ising machines. One important metric in this context is graph density, which is defined as the ratio of the number of edges \(E\) in the graph to the maximum possible number of edges in a graph with the same number of vertices \(V\). For  undirected graphs we consider in this paper, \(D = 2E / (V(V-1))\).  

\begin{figure*}[t!]
    \centering
    \includegraphics[width=1\linewidth]{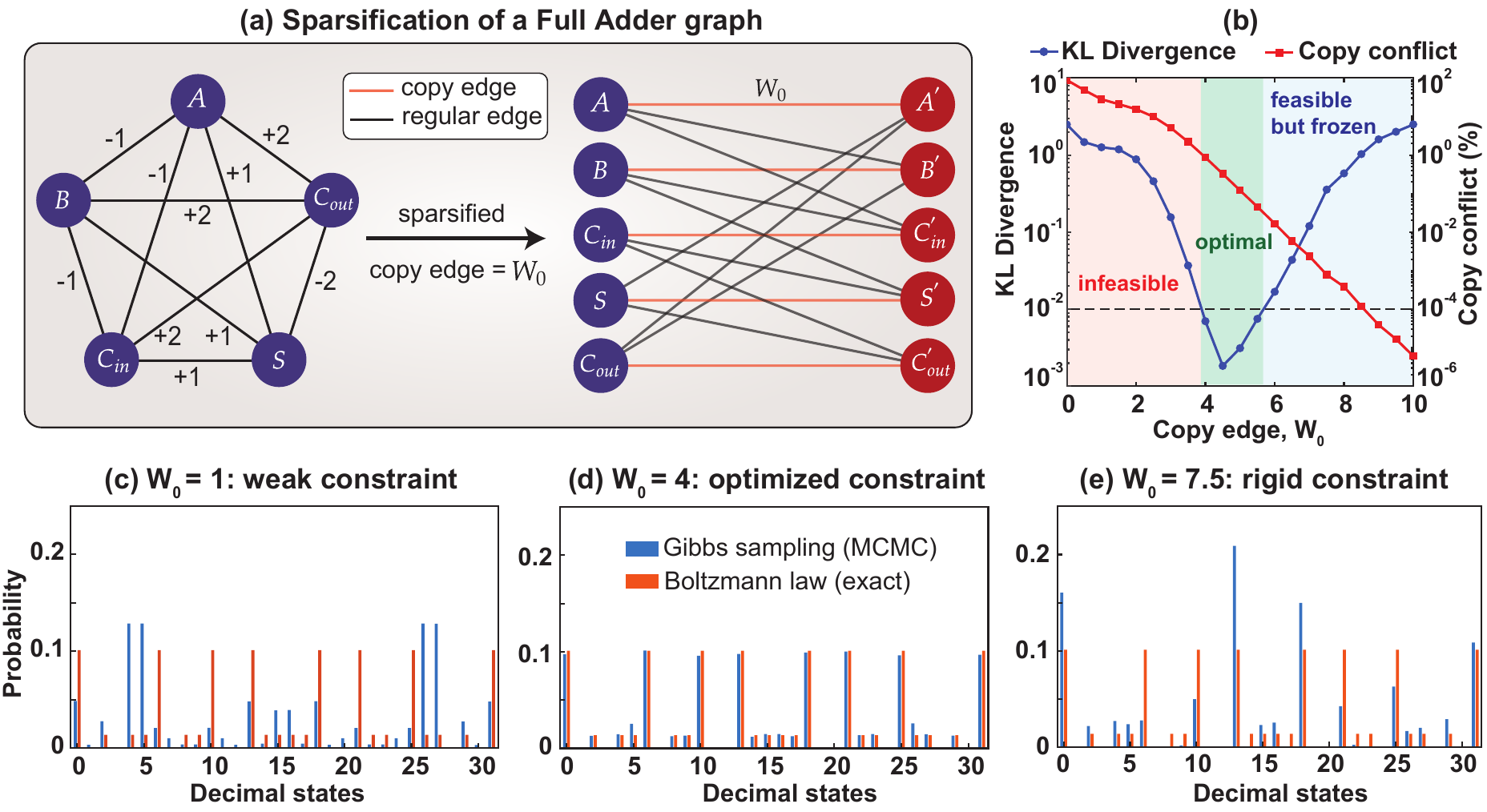}
    \vspace{-8pt}
    \caption{Sparsification of a full adder (FA). (a) A 5 p-bit all-to-all full adder is sparsified into a 10 p-bit network where $k$ is limited to 3, using 2 copies per node with copy edge $W_0$. (b) Kullback-Leibler divergence and copy conflict for the sparsified full adder example are computed from $10^7$ Monte Carlo sweeps and averaged over 20 independent Markov chains, indicating an optimal region near $W_0\approx 4.5$. (c) Comparison between the exact Boltzmann distribution and the experimental distribution obtained on the sparse graph with Gibbs sampling at $\beta =1$ and $10^{7}$ Monte Carlo sweeps. The experimental distribution is reduced to the original 32 states by resolving the copies as single spins if they agree or by doing coin flips if they disagree. When $W_0$ is weak, copies can take different values, e.g., $A\neq A'$, and the interpreted state is wrong. This is visible in the experimental distribution that does not match the Boltzmann distribution for the original graph. (d) Increasing $W_0$ to an optimized value, the copy nodes are in agreement, and the sparse network approximates the true Boltzmann distribution within the given $10^7$ Monte Carlo sweeps. (e) Having a larger $W_0$ ensures copy constraints are satisfied, but the rigidity of the coupling gets the network stuck in a few feasible states. Matching Boltzmann is guaranteed with more Monte Carlo sweeps, since the mapping is exact; however, this may not be feasible in practice.}
    \label{fig2}
    \vspace{-10pt}
\end{figure*}

The all-to-all graph in FIG.~\ref{fig1}a has $D$ = 100\% graph density. To implement Gibbs sampling (FIG.~\ref{fig1}c) in such a dense architecture, we need adders whose space complexity grows as $\mathcal{O}(N)$, since each p-bit needs a summation over all neighbors. On the other hand, FIG.~\ref{fig1}b shows the graph of a sparse Ising machine where each p-bit is connected to a predetermined and fixed number of neighbors,  \(k\) (\(k \ll N\)). As a result, the adder complexity is $\mathcal{O}(1)$, since each p-bit needs a summation of a fixed number of neighbors, independent of graph size $N$ as shown in FIG.~\ref{fig1}d. In frequency scaling, all-to-all connectivity faces a two-fold penalty: (1) since the p-bit adders grow linearly with $N$, they slow down with larger delay, and (2) due to the serial nature of Gibbs sampling as shown in FIG.~\ref{fig1}c (Algorithm 1), p-bits slow down linearly with $N$. Experimental results confirm this quadratic frequency drop as a function of network size, $\mathcal{O}(1/N^2)$. In contrast, sparse graphs enable parallel updates of independent p-bits. This can be achieved by coloring the sparse graph such that connected p-bits have different colors. All the colored p-bit blocks can then be updated in a single clock cycle using phase-shifted clocks assigned to each color \cite{aadit2022massively}. The fixed adder complexity combined with parallel p-bit updates keeps the sweep frequency constant as $\mathcal{O}(1)$ for sparse graphs. In practice, the frequency slightly varies due to growing clock trees and repeaters as we discuss in Section~\ref{sec:ASIC}. Digital synthesis results based on ASAP7 we show later confirm these scaling laws (Table~\ref{tab:ASIC-table}). 

It is important to distinguish that the sparse graphs shown in FIG.~\ref{fig1}(d-e) are not the result of sparsification of a dense problem, but rather represent {natively sparse} networks where each node has a fixed number of neighbors $k$, independent of system size $N$. These results serve to stress the fundamental architectural and timing advantages of sparse connectivity for Ising machines, namely, constant-time local summation and sweep frequency scaling. In later sections (e.g., FIG.~\ref{fig2} onward), we explore sparsification strategies to transform dense problems into such sparse forms, but the architectural arguments in FIG.~\ref{fig1} apply more broadly to any system that maintains fixed node degree, whether natively or through embedding.

\section{Sparsifying All-to-All Graphs}
\begin{figure}[t]
\setcounter{algorithm}{1} 
\begin{algorithm}[H]
\caption{Graph sparsification}
\label{alg:sparsification}
\begin{algorithmic}[1]
    \STATE \textbf{Input}: All-to-all matrix \( J_A \), copy edge \( W_0 \), maximum number of neighbors \( k \)
    \STATE \textbf{Output}: Sparsified matrix \( J_S \), copy indices for each node
    \STATE \( N \gets \text{number of nodes in } J_A \)
    \STATE \( J_S \gets J_A \)
    \STATE \( copies \gets \sim \max(\text{degree}(J_A)) / k \)
    \STATE \( indices  \gets \{\} \)
    \STATE \( copy \gets N + 1 \)
    \FOR{each node \( i \) in \( J_A \)}
        \STATE \( source \gets i \)
        \FOR{each of the \( copies \)}
            \STATE \( J_S(source, copy) \gets W_0 \)
            \STATE \( J_S(copy, source) \gets W_0 \)
            \STATE Move \( k - 1 \) edges of \( i \) to \( copy \) in \( J_S \)
            \STATE Append \( copy \) to \( indices[i] \)
            \STATE \( source \gets copy \)
            \STATE \( copy \gets copy + 1 \)
        \ENDFOR
    \ENDFOR
\end{algorithmic}
\end{algorithm}
\vspace{-10pt}
\end{figure}

Algorithm \ref{alg:sparsification} introduces copy nodes to limit the node degree \(k \ll N\) in an all-to-all graph with adjacency matrix \(\mathbf{J}_{\mathbf{A}}\). The number of copies required per node is given by the maximum initial degree divided by $k$. In practice, one must consider the parity of the maximum degree and $k$ to compute its exact value, which we omit here for simplicity. Then, for each node $i$ of the all-to-all matrix $J_A$ a ferromagnetic copy edge $W_0$ is inserted between the copy and the source (initially $i$) nodes to the extended sparse matrix $J_S$, while distributing some of the edges from $i$ to the new copy (no more than $k-1$). For each node $i$, a list keeps track of the copy indices, later used to decode the sparse graph back to the original one.

\begin{figure*}[t!]
    \centering
    \includegraphics[width=0.8\linewidth]{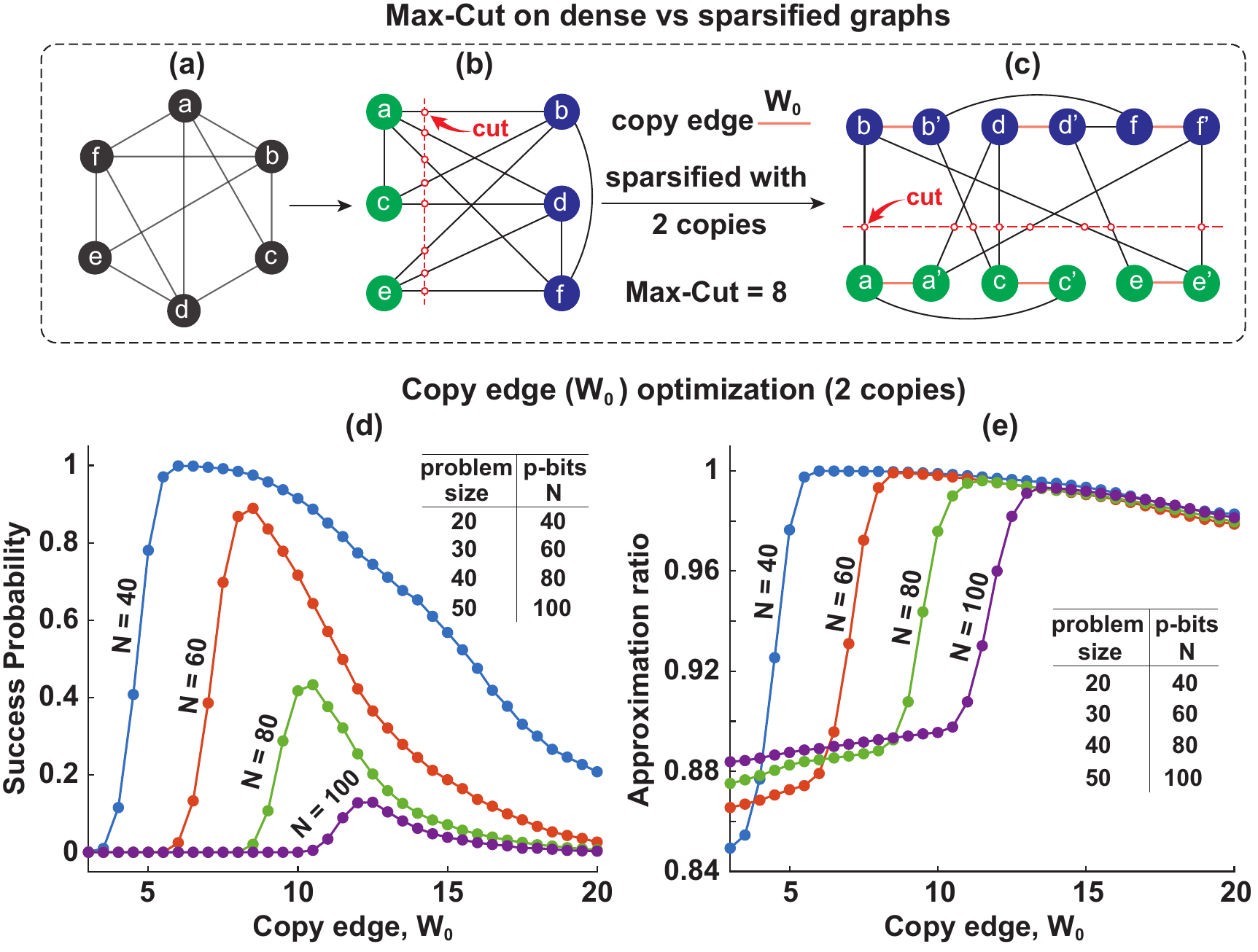}   
    \caption{(a) An arbitrary 6-node dense graph is assembled as (b) a bipartite graph to estimate the maximum cut. (c) Illustration of the dense Max-Cut instance sparsified with two copies per node to limit the node connectivity to $k$ = 3. Importantly, sparsification does not change the optimal cut when the copies agree. (d) Success probability for finding the Max-Cut as a function of copy edge $W_0$ for varying $N$. The Max-Cut instances have graph density = 0.75.
    For optimal success probability, $W_0$ needs to be carefully tuned for every size $N$  for the chosen Monte Carlo sweeps ($8 \times 10^5$ for all examples in these plots). Data points are averaged over 100 random instances and 100 trials per instance, each trial taking the mentioned Monte Carlo sweeps (MCS). Unbiased coin flips resolve copy spin conflicts. (e) We define an approximation ratio as (measured cut/optimal cut) and observe that for $8 \times 10^5$ MCS, the approximation ratio degrades more gracefully compared to success probability as a function of $W_0$. }
    \label{fig3}
\end{figure*}

Our sparsification method is similar in spirit to the minor graph embedding (MGE) technique used for quantum annealers \cite{choi2008minor, choi2011minor}. MGE needs to satisfy constraints on a predetermined target graph with an unspecified number of copy nodes. In practice, an embedding is often not found \cite{aadit2022massively}. The key difference of our approach is that MGE fixes the \textit{graph} while we fix the number of maximum \textit{neighbors} for a given node.  The MGE method is more restrictive due to the difficulties of building programmable superconducting circuits in different topologies.

For classical Ising machines, such as those implemented in FPGAs and ASICs, there is much greater flexibility, but as discussed in FIG.~\ref{fig1}, the maximum number of neighbors for a given node is the key metric to be minimized. Importantly, sparse graphs may not always need to be separately synthesized; instead, multiplexed, master-graph approaches for sparse graphs have been implemented in FPGAs \cite{nikhar2024all}.

However, our sparsification algorithm based on copy gates inherits a key difficulty of MGE, namely, the need to optimize the strength of the copy edge, $W_0$
\cite{hamerly_2019, venturelli_2015, pelofske2023comparing, willsch2022benchmarking, jain2021solving, le2023benchmarking, hayun2024determination, grant2022benchmarking}. FIG.~\ref{fig2} illustrates these points in a simple example. We start with a 5 p-bit all-to-all network (FIG.~\ref{fig2}a), whose low energy states correspond to the truth table of a Full Adder (FIG.~\ref{fig2}b). The FA graph is sparsified with 2 copies per p-bit to make it a 10 p-bit sparse network with a fixed $k$ = 3. In FIG.~\ref{fig2}c-e, we study the effect of copy gate strength on network dynamics. Note that FIG.~\ref{fig2}c-e correspond to a \textit{reduced} histogram for the 10 p-bit network where copies either agree or we do  coin flips to resolve the p-bit value if they disagree. When \(W_{0} = 1\) the constraints are too weak and the  copied nodes do not follow each other. The result is a poor match to the Boltzmann distribution. At the other extreme, for \(W_{0} = 7.5\) (rigid constraint), the copy chain enforces a very strong coupling between copies. The visited states are correct (obeying the constraint), but the system gets stuck due to the large coupling between copy gates, reminiscent of symmetry breaking in physics. Even after $10^7$ Monte Carlo sweeps, the true Boltzmann law is not recovered. The optimal balance for the chosen $10^7$ sweeps is achieved when \(W_{0} = 4\), where the copy edge strength enables a close match between the Gibbs sampling and the Boltzmann distributions. 

This example shows a fundamental difficulty arising in any ferromagnetic sparsification method (MGE or our method): the constraints are either weak, leading to ``chain breaking'' \cite{hamerly_2019, venturelli_2015}, or they are too rigid, leading to suboptimal searches. The necessity to optimize the edge strength for a \textit{given} number of Monte Carlo sweeps poses practical difficulties, as we discuss next.

\section{sparsification of dense Max-Cut}
Computing the maximum cut of a graph (Max-Cut) consists of finding two subsets of vertices $m_i=\pm1$ such that the number of edges between the two subsets is maximum. With weighted edges $J_{ij}$, the Max-Cut problem is expressed in the form of Eq.~\ref{eq:ising_hamiltonian} as:
\begin{equation}
    \text{Max-Cut}=\max_m \sum_{i<j} J_{ij}\frac{m_im_j-1}{2}
\end{equation}
 For our experiments, we generate random instances with 75\% edge density and binary weights for all sizes. The optimal cut is computed with the exact solver BiqCrunch \cite{BiqCrunch}. A 6-node dense Max-Cut instance example with optimal cut = 8 is depicted in FIG.~\ref{fig3}a, which is sparsified by introducing two copies per node, limiting the maximum neighbors to \(k = 3\). Ideally, if the copied nodes always agree with each other and the ground state is found, one can retrieve the original optimal cut (= 8). This example illustrates the exact equivalence of the original and the sparse graphs, showing how sparsification does not change the optimal cut for the original problem.

As discussed earlier, the copy edge  $W_{0}$ optimization is critical to find the optimal solution. We systematically show the optimization procedure for $W_{0}$  in FIG.~\ref{fig3}b  for sparsified graphs of Max-Cut instances (75\% initial density) of varying sizes. To obtain the results in FIG.~\ref{fig3}, we performed linear simulated annealing with a schedule of  $\beta = 0.125$ to  1 in steps of 0.125 with a total anneal time of 8$\times 10^5$ Monte Carlo sweeps. The final solution is estimated from the minimum energy state of the sparsified graph, obtained from the last 100 sweep readouts at the final $\beta$. Copy nodes are resolved using unbiased coin flips in case they disagree.

The initial dense graph sizes are $N$ = 20, 30, 40, and 50, and the sparse graph sizes become $N$ = 40, 60, 80, 100 respectively with two copies per node. The left plot shows the success probability of finding the Max-Cut as a function of \(W_{0}\). For different $N$, the peak success probability occurs at different $W_0$ for the anneal time, requiring a separate $W_0$  at each size. The peak also shifts towards higher values of $W_0$ as $N$ increases, indicating that the larger sizes require stronger copy edges to follow the increasing number of neighbors. These results are in qualitative agreement with those obtained from studies in quantum annealers \cite{venturelli_2015}. The results on the success probability of sparsified Max-Cut problems paint a dire picture: finding the optimal cut out of a given number of trials rapidly decays at a \textit{fixed} size. Note that the decay of success probability with increasing $N$ is expected, since the problem is NP-hard. 

\begin{figure*}[t!]
    \vspace{-5pt}
    \centering
    \includegraphics[width=1\linewidth]{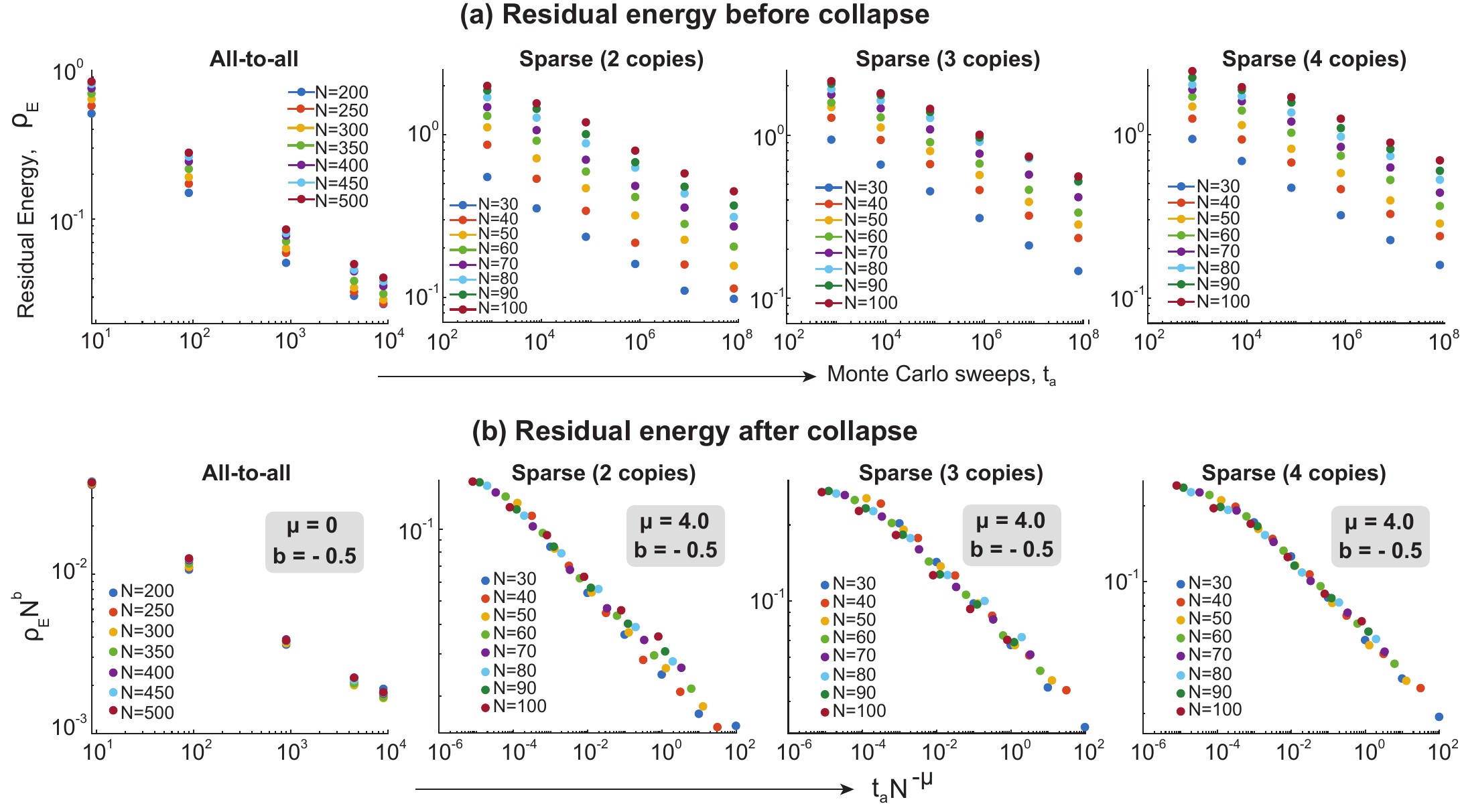}
    \vspace{-5pt}
    \caption{Residual energy as a function of Monte Carlo sweeps ($t_a$) plotted in log-log axes for (a) all-to-all and sparse configurations with 2, 3, and 4 copies for varying problem sizes. Data points are averaged over 60 and 20 Max-Cut instances (75\% density) for all-to-all and sparse graphs, respectively, with 50 trials per instance. Unbiased coin flips are used for sparse 2 copies, and majority votes for more than 2 copies. Larger problem sizes are chosen for all-to-all to observe the power law pattern, because smaller all-to-all problems trivially reach the ground state at this Monte Carlo sweep ($t_a$) range. Residual energy is defined as $\rho_{E}$ = (measured energy - ground energy)/$N$. (b) All of the raw curves for different $N$ fall onto a single line once the residual energy is multiplied by $N^{b}$ and the Monte Carlo sweeps ($t_a$) by $N^{-\mu}$, demonstrating finite-size scaling collapse with $\rho_{E}(N,t_a)\approx N^{b}F(t_a N^{-\mu})$.}
    \label{fig4}
    \vspace{-10pt}
\end{figure*}

However, in many practical applications, \textit{approximate solutions} close to optimum are often acceptable. Indeed, by their very nature, Ising machines are heuristic solvers and cannot be expected to reach the ground state of any hard problem with certainty. To investigate the effect of sparsification on \textit{approximation}, we define the approximation ratio metric, which is defined as the measured cut/optimal cut. 
FIG.~\ref{fig3}c shows the approximation ratio for sparsified graphs as a function of $W_0$. Strikingly, unlike the rapidly decaying success probability, the approximation ratio degrades much more gracefully over a very large range of $W_0$. It appears that reaching the \textit{optimal} cut seems to critically depend on the choice of $W_0$ whereas approximating it may not be as sensitive. In practice, in contexts where approximate optimization is acceptable, sparsification may lead to satisfactory results.
\vspace{-10pt}
\section{Finite-Size Scaling Analysis of Sparsification Overhead}
\label{sec:fss}
We analyze the residual energy per spin,
\[
\rho_E(N, t_a) = \frac{E(t_a) - E_{\mathrm{gs}}}{N}
\]
where \(E(t_a)\) is the energy after \(t_a\) Monte Carlo sweeps and \(E_{\mathrm{gs}}\) is the known ground state energy. One Monte Carlo sweep (MCS) is defined as a complete update of all spins in the network.

To compare performance across sizes and topologies, we adopt the finite-size scaling ansatz:
\begin{equation}
    \rho_E(N, t_a)N^b \approx F(t_a N^{-\mu}),
    \label{eq:fss}
\end{equation}
where \(b\) characterizes the scaling of residual energy and \(\mu\) captures the algorithmic slowdown due to sparsification. 

To extract the scaling exponents, we used \texttt{autoScale.py}~\cite{melchert2009autoscale}, an open-source tool for automated finite-size scaling analysis. We supply residual energy data at multiple system sizes and sweep times and optimize the rescaling parameters to achieve the best collapse. Although many combinations of \((b, \mu)\) can empirically yield good collapse over limited ranges, we follow theoretical predictions from theory to fix \(b\) and then determine the dynamic exponent \(\mu\) numerically. This approach allows us to interpret \(\mu\) directly as an overhead exponent attributable to sparsification.

We fix \(b = -\tfrac{1}{2}\), guided by theoretical results from Dembo, Montanari and Sen ~\cite{dembo2017extremal}. For dense Erdős–Rényi graphs \(G(N,p)\), the expected Max-Cut value scales as
\begin{equation}
    \mathrm{MaxCut}(G) = \frac{p}{4}N^2 + P^* \sqrt{\frac{p}{4}}\,N^{3/2} + \mathcal{O}(N^{3/2})
    \label{eq:maxcut_scaling}
\end{equation}
where \(P^* \approx 0.7632\) is the Parisi constant. The leading term corresponds to a random cut, while the subleading \(N^{3/2}\) correction lifts the solution above this baseline.

Since energy and cut value are related up to constants, the residual energy inherits the same finite-size structure as \(\mathrm{MaxCut}(G)\). In particular, the \(\mathcal{O}(N^{3/2})\) fluctuations in the cut translate into an \(\mathcal{O}(N^{1/2})\) scaling in the residual energy per spin:
\[
\rho_E(N) = \frac{E - E_{\text{gs}}}{N} \sim \mathcal{O}(N^{1/2})  \Rightarrow  b = -\tfrac{1}{2}
\]

This choice is valid under the assumption that the solver reaches energies within \( \mathcal{O}(N^{3/2}) \) of the optimal energy. If the algorithm stalls earlier (e.g., at a gap of \( \sim N^{3/2+\delta} \)), the scaling collapses may degrade and the choice of \(b = -\tfrac{1}{2}\) would no longer flatten the data. We observe that our solver consistently reaches energies within \(\mathcal{O}(N^{3/2})\) of the theoretical (extensive) ground state, supporting the \(b = -\tfrac{1}{2}\) assumption.

Finally, while sparsified networks reduce the number of neighbors per physical node (e.g., to \(\sim 0.375 N\) in our 2-copy, \(p = 0.75\) construction), the graph remains effectively dense, in the thermodynamic sense. Although sparsified graphs introduce \(\mathcal{O}(N)\) additional ferromagnetic couplings between copies, these edges are not part of the original logical problem. The energy $E(t_a)$ is computed after reducing the full sparse graph to its equivalent original logical graph using coin flips or majority votes. However, the ground state energy $E_{\mathrm{gs}}$ is computed from the original logical graph, and the number of logical spins normalizes residual energy. At the end of annealing, the copy couplings are strong enough to suppress chain breaks, so all auxiliary constraints are satisfied and do not introduce additive errors. As a result, the residual energy is governed entirely by the logical graph structure, and the universal exponent \(b = -\tfrac{1}{2}\) is assumed to be valid for sparsified instances.

To evaluate the scaling collapse, we generated dense Max-Cut instances with 75\% edge density. For each system size, we averaged the residual energy over 20 independently sampled graph instances, each with 50 independent Monte Carlo runs. In the all-to-all case, we used problem sizes up to \(N = 500\), limiting the maximum annealing time to \(t_a \leq 10^4\) sweeps so that finite-size effects are still visible. For sparsified graphs (2, 3, and 4 copies), we used logical system sizes from \(N = 30\) to \(100\).

Having justified our choice \( b = -\tfrac{1}{2} \), we now interpret the dynamic exponent \(\mu\) in Eq.~\ref{eq:fss} as a measure of the algorithmic time overhead required to reach a fixed residual energy. Specifically, \(\mu\) characterizes how the number of Monte Carlo sweeps \(t_a\) must scale with problem size \(N\) to achieve the same convergence behavior:
\[
\rho_E(N, t_a) N^{-1/2} \sim  F(t_a N^{-\mu})  \Rightarrow  t_a^{\text{sparse}} \sim N^{\mu} t_a^{\text{all-to-all}}
\]

In our experiments, we find that the all-to-all topology exhibits \(\mu \approx 0\), indicating fast convergence independent of system size. This is consistent with the observation that dense Erdős–Rényi Max-Cut graphs, despite being NP-hard, behave as mean-field systems under simulated annealing and are relatively easy to solve at moderate sizes. The convergence in this case is dominated by extensive contributions to the energy, and the algorithm is able to resolve the subleading \(\mathcal{O}(N^{3/2})\) fluctuations with minimal time overhead.

In contrast, sparsified networks exhibit \(\mu \approx 4\) across all tested copy counts (2, 3, and 4). While this indicates a polynomial slowdown in convergence, the fact that \(\mu\) remains approximately constant across increasing copy numbers is interesting. It suggests that sparsification introduces a fixed polynomial overhead rather than a runaway complexity. 

 The origin of a  consistent $\mu \approx 4$ overhead for all sparsified cases remains an open question. One possibility is that sparsification introduces two distinct sources of slowdown: local interactions restrict the propagation of information across the graph, and the introduction of copy chains adds internal delays in flipping logical spins. Although we measure time in Monte Carlo sweeps, where each spin in the physical graph is updated once per sweep, these effects may compound to stretch the convergence time in ways not present in the all-to-all setting. The observed scaling may reflect the interplay between these constraints, though we caution that this interpretation is speculative. Future work using more advanced Monte Carlo techniques such as parallel tempering may help reduce $\mu$ in practice. 

Even though sparsification introduces a steep polynomial overhead, all-to-all networks may ultimately not be realizable in hardware at scale   due to routing and fan-in constraints. In contrast, sparsified graphs may enable physically realizable and modular architectures with localized connectivity. 

Thus, our results illustrate a fundamental trade-off: sparsification introduces a constant polynomial time penalty (as captured by \(\mu\)), but enables constant-time sweep execution and compact area scaling in physical implementations. We also note that the overhead we observed may be problem dependent: the very low dynamic exponent of the all-to-all (dense) maxcut instances may be due to the mean-field nature of the problem. In truly frustrated spin glass instances where dynamic exponents are already high, the sparsification overhead may not be as steep, and our empirical experience sparsifying Circuit SAT instances supports this observation \cite{aadit2021computing}.

\begin{figure}[b!]
    \vspace{-5pt}
    \centering
    \includegraphics[width=0.99\linewidth]{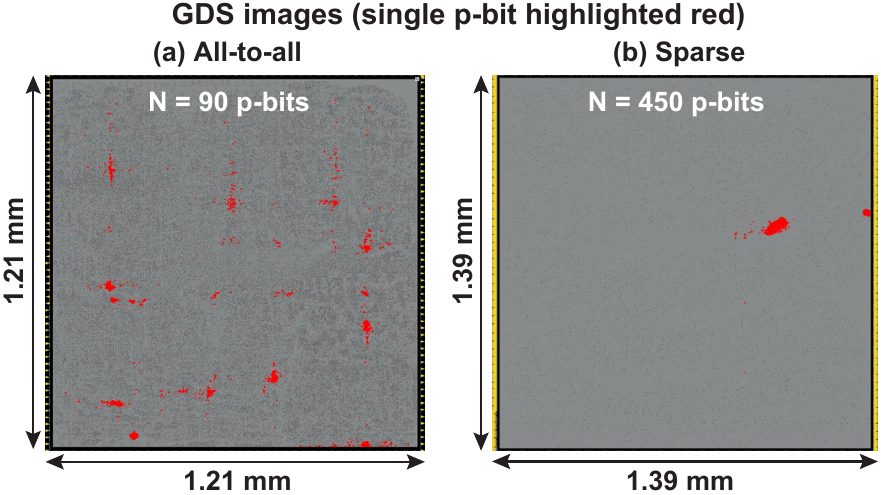}
    \caption{ (a) GDS images and chip dimensions of 90 p-bits with all-to-all connectivity. (b) Sparsifying the 90 p-bit all-to-all network into 450 p-bits (5 copies per p-bit) highlights design differences. A single p-bit, including its neighbor weights, activation lookup table, and pseudorandom number generator, is shown in red for both cases. In sparse graphs, the p-bit is localized, whereas in all-to-all, it is highly delocalized. Despite a 5X increase in network size, the chip area grew by only 1.3X. See \cite{aadit2022massively} for detailed p-bit RTL implementation.} 
    \label{figGDS}
    \vspace{-10pt}
\end{figure}  

\begin{figure}[t!]
    \centering
    \includegraphics[width=0.6555\linewidth]{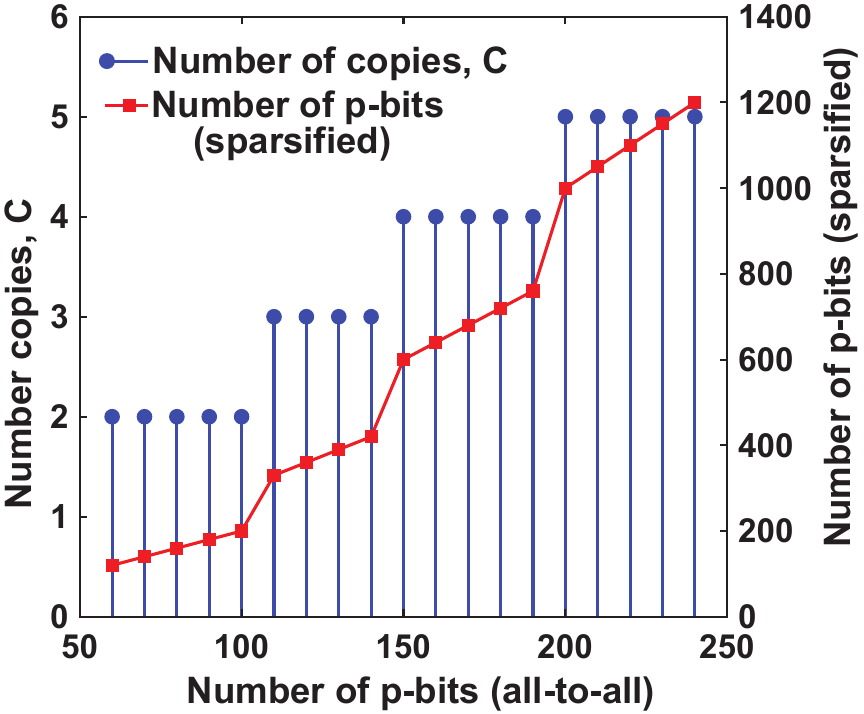} 
    \caption{All-to-all graphs are sparsified with a maximum allowed number of neighbors, $k = 51$. Consequently, the number of required copies, $C$, varies with $N$ as a staircase pattern where $C$ is the ceiling of $N/k$.}
    \label{fig6}
    \vspace{-10pt}
\end{figure}

\section{physical design considerations}
\label{sec:ASIC}
We now discuss physical design considerations of all-to-all vs. sparse network topologies. Our approach is based on register-transfer level (RTL) descriptions of our FPGA implementations and the ASAP7 process design kit \cite{clark2016asap7}, synthesized into a gate-level netlist using Genus. The flow proceeds with floorplanning and power planning, placement and global routing, clock tree synthesis (CTS), detailed routing, and the sign-off steps in {Innovus}. These steps ensure that the physical design is optimized for both area and performance while meeting the constraints imposed by the ASAP7 technology.

The GDS (Graphic Database System II) visualizations in FIG. \ref{figGDS} illustrates the physical implementation of the chips for (a) all-to-all and (b) sparse configurations, with the physical location of one random p-bit highlighted in red on the chips. The adders associated with each p-bit are spread across the chip to accommodate the extensive routing requirements in the all-to-all configuration. On the other hand, in the sparse configuration, adders are localized, reducing the overall routing complexity. Remarkably, despite growing 5X in network size for sparsification with 5 copies, the sparse chip only takes 1.3X the area of an all-to-all chip.

\begin{table}[b]
\vspace{-10pt}
\centering
\caption{area and frequency scaling of all-to-all vs. sparse Ising machines from physical design for fixed neighbors k=51}
\label{tab:ASIC-table}
\renewcommand{\arraystretch}{1.2} 
\resizebox{\columnwidth}{!}{%
\begin{tabular}{|c|cccccccc|}
\hline
\begin{tabular}[c]{@{}c@{}}\textbf{Problem}\\ \textbf{size}\end{tabular}& & 70  & 80  & 90 & 100 & 110 & 120 & 130 \\ \hline
\multirow{2}{*}{\textbf{p-bits}}& all-to-all & 70  & 80  & 90    & 100 & 110 & 120 & 130 \\
 & sparse & 140   & 160  & 180   & 200   & 330     & 360 & 390 \\ \hline
\multirow{2}{*}{\begin{tabular}[c]{@{}c@{}}\textbf{Frequency}\\ \textbf{(MHz)}\end{tabular}} & all-to-all & 711 & 672 & 621   & 583 & 522 & 529 & 471 \\
 & sparse & 1,060 & 1,033 & 1,022 & 1,010 & 1,021 & 1,004 & 1,012 \\ \hline
\multirow{2}{*}{\begin{tabular}[c]{@{}c@{}}\textbf{Sweep Time}\\ \textbf{(ns)}\end{tabular}} 
 & all-to-all & 98.5 & 119 & 145 & 172 & 211 & 227 & 276 \\
 & sparse & 3.77 & 3.87 & 3.91 & 3.96 & 3.92 & 3.98 & 3.95 \\ \hline
\multirow{2}{*}{\begin{tabular}[c]{@{}c@{}}\textbf{Total area}\\ \textbf{(mm²)}\end{tabular}}      & all-to-all & 0.627 & 0.813 & 1.023 & 1.268 & 1.529 & 1.818 & 2.241 \\
 & sparse & 0.721   & 0.890 & 1.092 & 1.314 & 1.895 & 2.260 & 2.586 \\ \hline
\multirow{2}{*}{\begin{tabular}[c]{@{}c@{}}\textbf{Area per p-bit}\\ \textbf{($\mu$m²/p-bit)}\end{tabular}} & all-to-all & 8,957 & 10,163 & 11,367 & 12,680 & 13,900 & 15,150 & 17,238 \\
 & sparse & 5,150 & 5,563 & 6,067 & 6,570 & 5,742 & 6,278 & 6,631 \\ \hline
\end{tabular}%
}
\vspace{-10pt}
\end{table}

To maintain $\mathcal{O}(1)$ scaling in sweep time, it is more appropriate to fix the maximum number of neighbors $k$ per p-bit rather than the number of copies $C$. With a fixed $k$, the required number of copies becomes a function of system size $N$, scaling approximately as $C \sim N/k$. This results in the staircase pattern shown in Fig.~\ref{fig6}, where $C$ increases in discrete steps as $N$ grows. In practice, architectural constraints (such as adder fan-in limits or routing complexity) determine the feasible value of $k$; for example, we used $k = 51$ in our design based on prior FPGA experience. For our experiments, we synthesized a single master sparse graph for each copy count $C = 2, 3,$ and $4$, allowing us to reuse these topologies to support a wide range of logical problem sizes in a consistent framework.

Table \ref{tab:ASIC-table} presents the performance metrics for two configurations: {all-to-all} and {sparse (with fixed neighbors, $k=51$)}. Synthesized (fully routed and signed off) results closely follow the theoretical expectations discussed in Fig.~\ref{fig1}d. We find that the sparse network area per p-bit grows slowly with $\mathcal{O}(N^{0.34})$ scaling. In contrast, the all-to-all network area per p-bit grows rapidly, showing $\mathcal{O}(N^{1.03})$ scaling. The sweep time trend also follows the theoretical expectations and FPGA-based experimental results in FIG.~\ref{fig1}e. The all-to-all sweep time increases following  $\mathcal{O}(N^{1.65})$ scaling, and the sparse network sweep time remains nearly constant with  $\mathcal{O}(N^{0.07})$.

 \begin{figure}[t!]
     \vspace{-5pt}
    \centering   
    \includegraphics[width=0.95\linewidth]{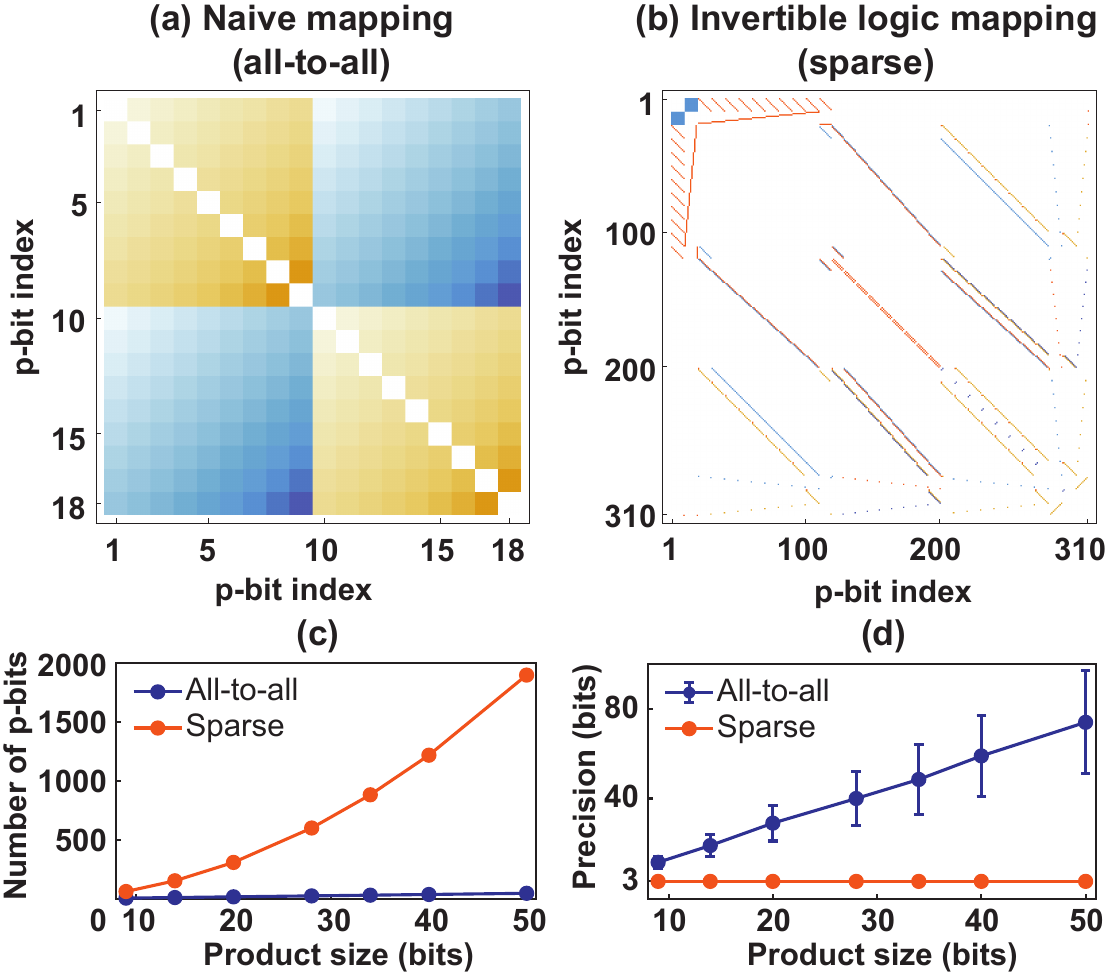}   
    \vspace{-5pt}
    \caption{Impact of problem mapping on sparsity: (a) Integer factorization can be mapped to Ising energy \(E=(F-pq)^2\), penalized when \(F \neq pq\). LSB of $p$ and $q$ are set to 1 since they are assumed to be odd. This naive mapping results in all-to-all connectivity with 18 p-bits, 100\% graph density, and high-order interactions (up to 4th order). An example with two 10-bit numbers (1007 and 1003) demonstrates the corresponding quadratic interaction matrix. (b) Mapping the problem to invertible logic gates produces a sparse network with 2.5\% graph density, requiring 310 p-bits. (c) The number of p-bits for invertible logic mapping increases more rapidly than the all-to-all p-bits as the problem size grows. (d) Dynamic weight range and precision comparison shows that all-to-all mapping demands significantly higher precision and weight range, while the sparse approach uses only four distinct weights with 3-bit precision. Error bars for all-to-all indicate minimum and maximum bit precision required for the weights.}
    \label{fig7}
    \vspace{-10pt}
\end{figure}

\section{Natively sparse formulations}
 An alternative to sparsification is to start from a natively sparse problem formulation.  For many Ising problems, there exist alternative native sparse representations.
 
 One way to see this is to consider the invertible logic formulation \cite{camsari2017stochasticL,onizawa2021sparse} of constraint satisfaction problems. Using principles of Boolean logic gates, Invertible logic can be used to construct circuits composed of p-AND, p-NOR and p-NOT gates that are probabilistic generalizations of ordinary AND, OR, NOT gates. The probabilistic formulation allows conditioning the outputs of composed logic gates so that the inputs are guided towards satisfying combinations. 

One instructive example is that of integer factorization.  This problem can be straightforwardly formulated as an optimization problem, similar to many of the dense formulations given by Ref.~\cite{lucas2014ising}, by defining the Ising energy as $E=(F-pq)^2$. Here,   $F$ is the semiprime and $p$, $q$ are the $n$-bit factors written in binary representation (the least significant bit is typically fixed to 1, since all prime numbers ($p,q$$>$2) are odd and this saves 2 bits). As shown in FIG.~\ref{fig7}a, this formulation results in an all-to-all graph, with nearly continuous weights. 

On the other hand, integer factorization can be elegantly expressed as an invertible multiplier circuit \cite{traversa2017polynomial,camsari2017stochasticL,aadit2022massively} built with probabilistic AND gates and full adders leading to graph with low density (FIG.~\ref{fig7}b).  As before, one drawback is that this representation requires more p-bits (FIG.~\ref{fig7}c), but it avoids the problem of starting from a dense network which may have to be aggressively sparsified with much greater overhead. The invertible multiplier technique has also shown superior performance in practice, by being able to factor very large semiprimes \cite{aadit2022massively} compared to all other Ising machines. The sparse mapping also benefits from requiring only four discrete integer weights, with 3-bit precision for \textit{any} product size, compared to the all-to-all mapping’s maximum 97-bit precision for the 50-bit factorization.

We presented integer factorization as a representative example of how different formulations can lead to networks with significantly different densities. Invertible logic can be used to directly represent a large class of constraint satisfaction problems, known as the Circuit SAT problem.

We also note that native representations may involve dense networks or more sophisticated generalizations such as the use of Potts spins to reduce transitions between invalid states~\cite{whitehead2023cmos} which can ultimately be more efficient for certain classes of optimization problems~\cite{iyer2025efficient}. Regardless, hardware limitations at extreme scales will still necessitate some form of sparsification or problem reduction for practical acceleration.

\section{Conclusion}
This work addresses the connectivity challenges of domain-specific Ising machines in solving optimization and sampling problems. Through FPGA-based p-bit Ising machines, we demonstrated that the scalability limitations of all-to-all networks make sparse alternatives essential. 
Sparsification especially when the starting network is dense comes with its own set of challenges: such as copy edge optimization, auxiliary nodes and increasing network sizes and increased time-to-solution. Despite these challenges, our results show that sparse Ising machines may deliver significant hardware advantages in area and frequency scaling.  ASIC-level designs corroborated these findings, highlighting the benefits of sparse networks for scalability \cite{srimani2024next}. Finally, we emphasized the importance of native sparse problem formulations as an alternative path to avoid dense graphs.

Another potential solution to sparsification can be obtained through the use of more sophisticated sampling algorithms. For instance, the parallel tempering algorithm \cite{replica_MC_1986,replica_exchange_1996} is known for efficiently finding ground states in complex energy landscapes and could help alleviate the rigidity imposed by the copy constraints $W_0$. Notably, a parallel tempering implementation with sparse replicas would still benefit from reduced sweep time, in contrast to all-to-all network frequencies that scale quadratically worse and become infeasible at large scales. Therefore, fast sparse hardware combined with advanced search algorithms presents a promising path toward large-scale Ising machines. These findings may be  broadly applicable for a general class of Ising machines. \vspace{15pt}

\section{Methods}
All of the experiments in this paper are performed in AMD Alveo U250, having Peripheral Component Interconnect Express (PCIe) connectivity. A fixed point precision of 10 bits (1 sign bit, 6 integer bits, and 3 fractional bits) is used for the weights ($J_{ij}$) modulated by the inverse temperature $\beta$. To sparsify a 100-node all-to-all graph, we introduced two, three, and four copies per node, resulting in sparse graphs with 200, 300, and 400 p-bits and corresponding average degrees of 51, 35, and 27, respectively. These serve as master graphs, which can be reconfigured to represent smaller sparse graphs: the 200-node graph supports problems with 2 copies and logical sizes from 40 to 100 nodes; the 300-node and 400-node graphs support 3-copy and 4-copy configurations for logical sizes from 60 to 100 and 80 to 100 nodes, respectively. This reuse allows a single FPGA implementation to accommodate a range of sparsified problem instances efficiently.

Max-cut instances are random Erdős–Rényi graphs with the probability of having an edge set to $p=0.75$. All the graph weights have values $W_{ij}=-J_{ij}=+1$. We compute the Max-cut values using the exact solver BiqCrunch \cite{BiqCrunch} for sizes $N\leq 100$. For larger sizes in the all-to-all analysis, we consider the best cut obtained with simulated annealing runs of $9 \times 10^4$ Monte Carlo sweeps.

\section*{Acknowledgment}
MMHS, NAA, KYC, and CD acknowledge support from the Office of Naval Research (ONR), Multidisciplinary University Research Initiative (MURI) grant N000142312708. NAA and KYC acknowledge support from the Semiconductor Research Corporation (SRC) grant. TW, CS, DC, AR, and TS acknowledge support from Samsung, Carnegie Mellon University Dean’s Fellowship and Tan Endowed Graduate Fellowship in Electrical and Computer Engineering, Carnegie Mellon University.

\end{document}